\documentclass[aps,preprint]{revtex4-1}
\usepackage[dvipdfmx]{graphicx,color}
\usepackage{bm}
\usepackage{amsmath,amsthm,amssymb}
\usepackage{mathrsfs}
\usepackage{gt15}
\usepackage{color}

\begin{document}

\title{
Effective Hamiltonian approach to optical activity in Weyl spin-orbit system} 

\author{Hideo Kawaguchi$^{1,2}$}
\email[Email address: ]{hideo.kawaguchi@riken.jp}
\author{Gen Tatara$^2$}
\email[Email address: ]{gen.tatara@riken.jp}
\affiliation{$^1$ Graduate School of Science and Engineering, Tokyo Metropolitan University, Hachioji, Tokyo 192-0397, Japan \\
$^2$ RIKEN Center for Emergent Matter Science (CEMS), 2-1 Hirosawa, Wako, Saitama 351-0198, Japan}

\date{\today}

\begin{abstract}
Chirality or handedness in condensed matter induces anomalous optical responses such as natural optical activity, rotation of the  plane of light polarization, as a result of breaking of spatial-inversion symmetry. 
In this study, optical properties of a Weyl spin-orbit system with quadratic dispersion, a typical chiral system invariant under time-reversal, are investigated theoretically by deriving an effective Hamiltonian based on an imaginary-time path-integral formalism. 
We show that the effective Hamiltonian can be indeed written in terms of an optical chirality order parameter suggested by Lipkin. The  natural optical activity is discussed based on the Hamiltonian.
\end{abstract}

%%%%%%%%%%%%%%%%%%%%%%%%%%%%%%%%%%%%% \textquotedblleft covariant\textquotedblright\ one,
\maketitle

\section{Introduction}  \label{I}
Spin-orbit interaction that couples electron motion to its spin as a consequence of a relativistic effect plays an important role of  mixing of electric and magnetic degrees of freedom. Although the spin-orbit interaction derived from the Dirac equation is weak in vacuum, it can be significantly large in solids without spatial-inversion symmetry realized in the interfaces and surfaces, when heavy elements are involved \cite{Ast07}. Of particular current interest is electromagnetic cross-correlation effects induced by strong Rashba interaction at interfaces \cite{Rashba60}, whose Hamiltonian is expressed as ${\cal{H}}_{\rm R} = - \alphaRv \cdot (\pv \times \bm{\sigma})$, where $\bm{p}$ is the linear momentum of the electron, $\bm{\sigma}$ is the vector of Pauli matrices representing the electron spin, and $\alphaRv$ is a polar axis, called the Rashba field, representing the inversion symmetry breaking. 
The Rashba interaction leads to generation of magnetization from an applied electric field (Edelstein effect \cite{Edelstein90}) and an electric current from a magnetic field or the magnetization (inverse Edelstein effect \cite{Shen14}). 
Experimentally, the magnetization by Edelstein effect is observed using Kerr effect at an interface between Cu, Ag and Bismuth oxide \cite{Puebla17}. The electric current induced by inverse Edelstein effect is measured in a multilayer of Ag, Bi and ferromagnet \cite{Sanchez13}. 

Those cross-correlation effects give rise to antisymmetric off-diagonal components in an electric permittivity tensor, resulting in an anomalous optical response for linearly-polarized waves such as birefringence and negative refraction
\cite{Shibata16, Shibata18}. 
Furthermore, in the case of broken time-reversal invariance due to magnetization or magnetic field, anisotropic light propagation (directional dichroism) independent of light polarization emerges because of Doppler shift induced inversion-odd diagonal components due to an intrinsic flow \cite{Kawaguchi16}. 

Optical responses and cross-correlation effects are qualitatively discussed based on symmetry argument. An optical response for circularly-polarized waves, natural optical activity, is caused by the inversion symmetry breaking which induces an antisymmetric off-diagonal component linear in the wave vector of light in an electric conductivity tensor \cite{Landau84}. 
Natural optical activity of chiral molecules was phenomenologically discussed in Ref. \cite{Condon37}. 
It was pointed out that chiral nature leads to the electric flux density and magnetic field strength, $\bm{D}$ and $\bm{H}$, given by 
\begin{align}
\bm{D}=&\epsilon\biggl[\Ev -\frac{g}{\epsilon} \dot{\bm{B}}\biggl] ,\nnr
\bm{H}=&\frac{1}{\mu}\biggl[\bm{B}-\mu g\dot{\Ev}\biggl], \label{ADH}
\end{align}
where $\Ev$ and $\bm{B}$ are electric and magnetic fields, respectively. $\epsilon$ and $\mu$ are the electric permittivity and magnetic permeability in the medium, respectively, and 
the constant $g$ characterizes the breaking of spatial inversion symmetry. 
It was demonstrated through discussion on a correlation function that $g$ is finite when the inversion symmetry is broken \cite{Nakano69}.

In the case of electron systems, a typical chiral system is the one with the Weyl spin-orbit interaction (Refs. \cite{Samokhin07, Mineev10, Mineev13}), whose Hamiltonian is given by 
\begin{align}
{\cal{H}}_{\rm W}=-\lambda(\bm{p}\cdot\bm{\sigma}), \label{Hw}
\end{align}
where $\lambda$ stands for the coupling constant of the spin-orbit interaction with broken inversion symmetry. 
One typical example is a metallic crystal Li$_{2}$(Pd$_{1-x}$,Pt$_{x}$)$_{3}$B \cite{Samokhin07}. 
The interaction, Eq. (\ref{Hw}), breaks mirror symmetry with respect to all the three axes (Ref. \cite{Ma15}), resulting in a radial electron spin texture on Fermi surface \cite{Yoda15}. The radial electron spin texture generates magnetization by the electric field in the system having helical structure such as Se or Te \cite{Yoda15}. In this paper, we examine natural optical activity in the Weyl spin-orbit system by deriving an effective Hamiltonian for electromagnetic fields. 
The Weyl system is shown to exhibit different optical responses from the Rashba case studied in Refs. \cite{Shibata16, Shibata18}, because the symmetry is different. 

The Weyl spin-orbit systems, where time-reversal symmetry is kept, are called truly chiral systems \cite{Barron86}. 
In contrast, quantity which breaks time-reversal invariance besides spatial one is called 
\textquotedblleft false chirality\textquotedblright\ . 
A symmetry analysis showed that the truly chiral interaction $\pv\cdot\bm{\sigma}$ leads to natural optical activity in chiral molecules \cite{Barron86}. In the case of electromagnetism, quantities with \textquotedblleft false chirality\textquotedblright\ are $\Ev\cdot\Bv$ and $\Ev\times\Bv$. 
The scaler product $\Ev\cdot\Bv$ appears in an effective Hamiltonian of 3+1-dimensional Weyl semimetal in the form  ${\cal{H}}_{\theta}=\theta(\rv, t)\Ev\cdot\Bv$ \cite{Zyuzin12, Goswami13}, where $\theta(\rv, t)$ is a topological field depending on space $\rv$ and time $t$. The effective Hamiltonian leads to topological electromagnetic cross-correlation effects such as chiral magnetic effect and anomalous Hall effect \cite{Vazifeh13}. The vector product form $\Ev\times\Bv$ appears in magnetic Rashba conductors \cite{Kawaguchi16} or insulator multiferroics \cite{Spaldin08}. The effective Hamiltonian in this case is ${\cal{H}}_{u}=\uv\cdot(\Ev\times\Bv)$ with a constant vector $\uv$ representing an intrinsic flow and it could cause directional dichroism \cite{Kawaguchi16}. 

%%%%%%%%%%%%%%%%%%%%%%%%%%%%%%%%%%%%%%%%%%%%%%%%%%%%%%%%%%%%%%%%%%%%%%%%%%%%%%%%%%%%%%%%%%%
\section{Phenomenological study}  \label{II}
Let us discuss the effective Hamiltonian for electromagnetic fields in the Weyl spin-orbit system from the symmetry point of view.

In vacuum, the effective Hamiltonian of electromagnetic field is \cite{Landau71}
\begin{align}
H=&\int {\rm d}^3r \frac{1}{2}\left(\epsilon_{0}|\Ev|^2+\frac{1}{\muz}|\Bv|^2\right), \label{EM}
\end{align}
where $\epsilon_{0}$ and $\muz$ are the electric permittivity and magnetic permeability of the vacuum. 
When coupled to electron system lacking spatial-inversion symmetry, an interaction linear both in $\Ev$ and $\Bv$ is expected to arise in the form 
\begin{align}
H_{EB}
=&g\int {\rm d}^3r (\Bv\cdot\dot{\Ev}-\Ev\cdot\dot{\Bv})\equiv \frac{2g}{\epsilon_{0}}C_{\chi}, \label{Hop}
\end{align}
where $g$ is a constant reflecting the breaking of spatial-inversion symmetry 
and $C_{\chi}$ is an optical chirality order parameter defined in Refs. \cite{Lipkin64, Proskurin17}.

Here, we show that the interaction of Eq. (\ref{Hop}) leads to optical activity. 
Maxwell's equations (equation of motion) including $H_{EB}$ are given by 
\begin{align}
 \bm{\nabla}\cdot\Ev =& \frac{\rho}{\ez}+ \frac{g}{\ez}\bm{\nabla}\cdot\dot{\Bv}, \nnr
 \bm{\nabla}\times\Bv =& \muz\jv+\ez\muz\frac{\partial \Ev}{\partial t} -\muz g\frac{\partial^2 \Bv}{\partial t^2}+\muz g\bm{\nabla}\times\dot{\Ev}, \label{Maxwell}
\end{align}
where $\rho$ represents the charge density and $\bm{j}$ is the charge-current density.
The condition of the absence of monopole and the Faraday's induction law are not changed because of U(1) gauge symmetry. From Eq. (\ref{Maxwell}), we obtain the electric and magnetic fields including the cross-correlation effect due to Eq. (\ref{Hop}) as 
\begin{align}
 \Ev_{\rm tot} \equiv& \Ev-\frac{g}{\ez}\dot{\Bv}, \nnr
 \Bv_{\rm tot} \equiv& \Bv-\muz g\dot{\Ev}. \label{EBcross}
\end{align}
The above expressions are indeed identical to Eq. (\ref{ADH}) suggested by Ref \cite{Condon37} in the context of the optical activity for circularly-polarized waves. 
\begin{figure}
\centering
\includegraphics[width=4.3cm]{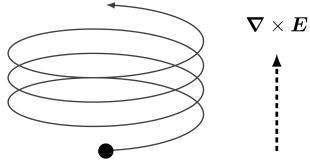}
\caption{Schematic illustration of charged particle's helical motion under the effect of $g\bm{\nabla}\times\Ev$. Filled circle and solid arrow stand for the particle and its orbital motion, respectively.}
\label{fig:helical}
\end{figure}
Using $\dot{\Bv}=-\bm{\nabla}\times\Ev$, $\Ev_{\rm tot}$ in Eq. (\ref{EBcross}) is rewritten as
\begin{align}
 \Ev_{\rm tot} =& \Ev+\frac{g}{\ez}\bm{\nabla}\times\Ev. \label{rot}
\end{align}
The equation (\ref{rot}) clearly describes a chiral nature of the system. In fact, it indicates that the electric field acquires an additional component proportional to its rotation, $\bm{\nabla}\times\Ev$. When a charge undergoes a circular motion by $\Ev(\rv)$, therefore, the motion is drifted in the perpendicular direction due to the term $g\bm{\nabla}\times\Ev$, resulting in a helical motion shown in Fig. \ref{fig:helical}. This helical motion gives rise to optical activities. 

In fact, we see directly that Eq. (\ref{EBcross}) results in the optical activity for circularly-polarized waves by deriving the dispersion relation of light. 
From Eq. (\ref{Maxwell}), the wave equation in the medium reads 
\begin{align}
\sum_{\nu}[c^2(\kv^2\delta_{\mu\nu}-k_{\mu}k_{\nu})-\omega^2\epsilon_{\mu\nu}]E_{\nu}=&0 , \label{wave}
\end{align}
where $c$ is the light velocity in vacuum, $\kv$ and $\omega$ are the wave vector and angular frequency of electromagnetic waves, respectively, and
\begin{align}
\epsilon_{\mu\nu}&\equiv\delta_{\mu\nu}
+i\frac{g}{\ez}\sum_{l}\epsilon_{\mu l\nu }k_{l},
\label{ep}
\end{align}
is an electric permittivity tensor having the antisymmetric off-diagonal component linear in $\kv$ due to the violation of spatial-inversion symmetry \cite{Baranova77, Train08}. Here $\epsilon_{\mu l \nu}$ is a totally antisymmetric tensor. 
Since Eqs. (\ref{wave}) and (\ref{ep}) give the characteristic equation for plane waves traveling along in $z$-axis of the form 
\begin{align}
\begin{vmatrix}
c^2\kv^2 -\omega^2 & -\omega^2\epsilon_{xy} & 0 \\ \omega^2\epsilon_{xy} & c^2\kv^2 -\omega^2 &0 \\
0 & 0 & -\omega^2
\end{vmatrix}
=0 ,
\end{align}
we obtain a dispersion relation,
\begin{align}
\kv^2 =& \frac{\omega^2}{c^2} \biggl[1 \pm \frac{g}{\ez}|\kv|
\biggl],
\end{align}
where $\pm$ stands for the sense of circular polarization. 
Therefore, the existence of the optical chirality order parameter leads to a rotation of the electric field in a plane perpendicular to the incident direction, namely, circular dichroism, as pointed out in Ref. \cite{Tang10}. 

Originally, the optical chirality order parameter was mathematically introduced to describe the solution of Maxwell's equations in order to explore conserved physical quantities reflecting the symmetry of electromagnetic fields \cite{Lipkin64}. Lipkin called it zilch, meaning that it has no physical effects. The quantity is revisited recently as it determines the polarization of circularly-polarized light \cite{Bliokh11, Bliokh15}. The cross-correlation effects was phenomenologically discussed in terms of the optical chirality in Ref. \cite{Proskurin17}. 
Until now, however, the optical chirality has not been discussed based on a microscopic ground. 
The aim of the present study is to show that the optical chirality indeed appears in the effective Hamiltonian by an imaginary-time path-integral formalism, and present a microscopic scenario on how the optical chirality leads to circular dichroism. 

%%%%%%%%%%%%%%%%%%%%%%%%%%%%%%%%%%%%%%%%%%%%%%%%%%%%%%%%%%%%%%%%%%%%%%%%%%%%%%%%%%%%%%%%%%%
\section{Derivation of effective Hamiltonian}
\begin{figure}
\centering
\includegraphics[width=8.6cm]{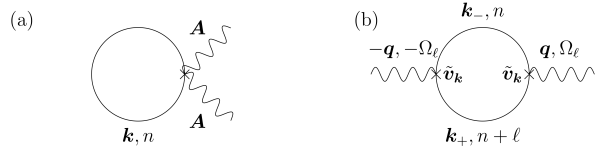}
\caption{Diagrammatic representation of the contribution to the effective Hamiltonian. Solid lines represent the thermal Green's function for electron and the wavy lines denote the gauge field, respectively. Diagrams, $(\rm a)$ and $(\rm b)$, correspond to the contributions of $n_{\rm e}$ and $\chi_{jj}^{\mu\nu}$ in Eq. (\ref{lnZ}), respectively.}
\label{fig:fd}
\end{figure}

In this section, we derive the effective Hamiltonian based on the imaginary-time path-integral formalism \cite{Sakita85}. 
We set $\hbar=1$ for simplicity, where $\hbar$ is the Planck constant divided by $2\pi$. 
We consider an electron system having the Weyl spin-orbit interaction under the effect of electromagnetic fields described by the gauge field $\Av$. In the field-representation, the conduction electrons are characterized by two-component annihilation and creation fields, ${c} (\bm{r},\tau)$ and $\bar{c} (\bm{r},\tau)$, with spin up and down along the $z$-axis, where ${c}$ and $\bar{c}$ are defined on an imaginary time $\tau$. The imaginary-time Lagrangian of the system thus reads $L[\bar{c},c,\bm{A}]= L_{\text 0} + L_{A}$, where 
\begin{align} 
L_{\text 0}(\tau)&\equiv \int {\rm{d}}^{3} r \bar{c}\biggl[ 
\frac{\partial}{\partial\tau} - \left(\frac{\bm{\nabla}^2} {2m} + \mu \right)
+\frac{i\lambda}{2}(\overleftrightarrow{\nabla} \cdot \bm{\sigma})\biggl]c, \\
L_{A}(\tau)&\equiv -\int {\rm{d}}^{3} r \bm{A}
\cdot \left(\frac{ie}{2m}\bar{c} \overleftrightarrow{\nabla}c  - \frac{e^2}{2m}  \bm{A}\bar{c}c+e\lambda\bar{c}\bm{\sigma}c\right).
\end{align}
Here $m$ is the electron mass, $\mu$ is the chemical potential of the system, $\lambda$ stands for the coupling constant of the Weyl spin-orbit interaction, $\bm{\sigma}$ is the vector of Pauli matrices, $\bar{c} \overleftrightarrow{\nabla} c $ $\equiv$ $\bar{c}\left(\bm{\nabla} c \right)-\left(\bm{\nabla} \bar{c} \right)c$, and $-e$ is the electron charge ($e>0$). 
The effective Hamiltonian for the electromagnetic field $H_{\rm eff}[\Av]$ is defined as
\begin{align}
\int_{0}^{\beta}{\rm d}\tau H_{\rm eff}[\Av]\equiv 
-{\rm ln}\mathcal{Z}[\Av]
, \label{HA}
\end{align}
where $\mathcal{Z}$ is the partition function in path-integral representation and $\beta$ denotes the inverse temperature. 
Equation (\ref{HA}) is calculated by integrating out the conduction electrons in the partition function as  
\begin{align}
\mathcal{Z}[\bm{A}] = \int \mathcal{D}\bar{c}\mathcal{D}ce^{-\int_0^\beta d\tau L[\bar{c},c,\bm{A}]},
\end{align}
where $\mathcal{D}$ stands for the path-integral. 
By carrying out the path integral over the electrons, the contribution to the second order in the gauge field reads (diagrammatically shown in Fig. \ref{fig:fd}) 
\begin{align}
\ln \mathcal{Z} &=
-\int_{0}^{\beta} {\rm{d}}\tau\int {\rm{d}}^3r \sum_{\mu\nu}A_{\mu}A_{\nu}\frac{e^2}{2m}n_{\rm e}(\bm{r},\tau)\delta_{\mu\nu} \nnr
&+\frac{1}{2}\int_{0}^{\beta} {\rm{d}}\tau\int_{0}^{\beta} {\rm{d}}\tau'\int {\rm{d}}^3r\int {\rm{d}}^3r' \sum_{\mu\nu} \nnr
&\times A_{\mu}A_{\nu}\chi_{jj}^{\mu\nu}(\bm{r},\bm{r}',\tau,\tau').  \label{lnZ}
\end{align}
Here $n_{\rm e}(\bm{r},\tau)\equiv \langle\bar{c}(\bm{r},\tau)c(\bm{r},\tau)\rangle$ is the electron density and 
\begin{align}
\chi_{jj}^{\mu\nu}(\bm{r},\bm{r}',\tau,\tau')&\equiv \langle \tilde{j}_{\mu}(\bm{r},\tau) \tilde{j}_{\nu}(\bm{r}',\tau') \rangle \label{xjj}
\end{align}
represents the current-current correlation function. The total electric current is denoted by   
\begin{align}
 \tilde{j}_{\mu}
 \equiv& -j_{\mu}+2e\lambda s_{\mu},
\end{align}
where $\bm{j}\equiv -\frac{ie} {2m}\bar{c}\overleftrightarrow{\nabla}c$ and $\bm{s}\equiv \frac{1}{2} \bar{c}\bm{\sigma}c$ are the bare electric current density and electron spin density, respectively. The thermal average $\langle\ \rangle$ in Eq. (\ref{xjj}) is calculated in the equilibrium state determined by the Lagrangian $L_{\text{0}}(\tau)$. 
Using Wick's theorem, the electron density and the correlation function are expressed as 
$n_{\rm e}=-\frac{1}{\beta V}\sum_{n,\kv}{\rm tr}[{\mathscr{G}}_{\kv,n}]$ and
\begin{align}
\chi_{jj}^{\mu\nu}(\qv,i\Omega_{\ell})&=-\frac{e^2}{\beta V}\sum_{n,\kv}{\rm tr} [\tilde{v}_{\kv,\mu}{\mathscr{G}}_{\kv_{+},n+\ell}\tilde{v}_{\kv,\nu}{\mathscr{G}}_{\kv_{-},n}], \label{chi} 
\end{align}
respectively, where $\tilde{\bm{v}}_{\kv}\equiv\bm{v}-\lambda \bm{\sigma}$ with $\bm{v}\equiv\frac{\kv}{m}$, ${\rm tr}$ is the trace over spin space, $V$ is the volume of the system, and 
\begin{align}
{\mathscr{G}}_{\kv,n}&\equiv 
\frac{1}{i\omega_{n}-\epsilon_{\kv}-\bm{\gamma}_{\kv}\cdot\bm{\sigma}+i\eta{\rm sgn}(\omega_n)}
\end{align}
is the $2\times 2$ thermal Green's function for electrons. It includes the Weyl spin-orbit interaction and a finite electron-elastic-scattering lifetime $\tau_{\rm e}$ as $\eta\equiv \frac{1}{2\tau_{\rm e}}$, and ${\rm sgn}(\omega_n)$ $\equiv$ $1$ and $-1$ for $\omega_n > 0$ and $\omega_n < 0$, respectively. 
Here $\kv_{\pm}\equiv\kv\pm\frac{\qv}{2}$, $\epsilon_{\kv}=\frac{k^2}{2m}-\mu$ is the electron energy measured from the Fermi energy, $\bm{\gamma}_{\kv}\equiv -\lambda \kv$, and $\kv$ and $\omega_n\equiv\frac{(2n+1)\pi}{\beta}$ with $n$ being an integer are the wave vector and fermionic thermal frequency of the conduction electron, respectively. The wave vector and thermal frequency carried by the gauge field are denoted by $\qv$ and $\Omega_{\ell}\equiv\frac{2\pi\ell}{\beta}$ with $\ell$ being an integer, respectively. 

Since we are interested in the effective Hamiltonian in the long-wavelength and low-energy region, we expand the correlation functions Eq. (\ref{chi}) with respect to $\qv$. Up to the first order of $\qv$, Eq. (\ref{chi}) reduces to (see Appendix\ref{A})
\begin{align}
\chi_{jj}^{\mu\nu}(\qv,i\Omega_{\ell})&\simeq \chi_{jj}^{\mu\nu}(\qv=0,i\Omega_{\ell})- ig(i\Omega_{\ell})\sum_{\rho}\epsilon_{\mu\rho\nu}q_{\rho}, \label{chiq}
\end{align}
with
\begin{align}
g(i\Omega_{\ell})&\equiv\frac{e^2\lambda^3}{24}\sum_{\kv}\sum_{\sigma_{1}\sigma_{2}\sigma_{3}}\xi_{\kv,\sigma_{1}\sigma_{2}\sigma_{3}} \nnr
&\times \left(-\frac{1}{\beta}\right)\sum_{n}{\rm g}_{\kv, n, \sigma_{1}}{\rm g}_{\kv, n, \sigma_{2}}({\rm g}_{\kv, n+\ell, \sigma_{3}}+{\rm g}_{\kv, n-\ell, \sigma_{3}}), \label{g}
\end{align}
where 
\begin{align}
\xi_{\kv, \sigma_{1}\sigma_{2}\sigma_{3}}\equiv
\frac{k}{\lambda m}(\sigma_{3}-3\sigma_{1}\sigma_{2}\sigma_{3} +2\sigma_{2}) 
+3-(\sigma_{1}\sigma_{2}+2\sigma_{2}\sigma_{3})
. \label{xi}
\end{align}
Here ${\rm g}_{\kv,n,\sigma}\equiv \left[i\omega_{n}-\epsilon_{\kv}^{\sigma}+i\eta{\rm sgn}(\omega_n)\right]^{-1}$ with  $\epsilon_{\kv}^{\sigma}=\epsilon_{\kv}+\sigma\lambda|\kv|$ is the Green's function diagonalized in the spin space and  $\sigma_{i}$ = $\pm 1$ ($i$=1$\sim$3) is the diagonalized spin index. 

$\chi_{jj}^{\mu\nu}(\qv=0)$ and the coefficient $g$ on the right-hand side of Eq. (\ref{chiq}) are calculated by the analytic continuation. 
We first show that the first term with $\qv=0$ is irrelevant. 
Expanding $\chi_{jj}^{\mu\nu}(\qv=0)$ with respect to the external frequency $\Omega$ defined by the analytic continuation to $\Omega +i\delta\equiv i\Omega_{\ell}$ \cite{ADG75,AS06}, where $\delta$ is a small positive imaginary part, the result up to the second order in $\Omega$ reduces to
\begin{align}
\chi_{jj}^{\mu\nu}(\qv=0,\Omega +i\delta)&\simeq \chi_{jj}^{\mu\nu,\Omega^{0}}+\chi_{jj}^{\mu\nu,\Omega^{1}}+\chi_{jj}^{\mu\nu,\Omega^{2}}
. \label{chiq0}
\end{align}
The first term in the above equation is $\chi_{jj}^{\mu\nu,\Omega^{0}}=-\frac{e^2}{m}n_{\rm e}\delta_{\mu\nu}$; hence, the contribution of $\chi_{jj}^{\mu\nu,\Omega^{0}}$ and that of $n_{\rm e}$ shown in Eq. (\ref{lnZ}) cancel each other, as is required by  the gauge invariance. The second term on the right-hand side of Eq. (\ref{chiq0}), $\chi_{jj}^{\mu\nu,\Omega^{1}}$, is a term with  $i\Omega\delta_{\mu\nu}$, but we drop the term proportional to $i\Omega A_{\mu}(-\Omega)A_{\mu}(\Omega)$ by noting the fact that $\int {\rm d}t A_{\mu}\dot{A}_{\mu}=0$. In Eq. (\ref{chiq0}), there appear the odd orders with respect to $\Omega$, but these terms also become total differential with respect to time. 
The third term on the right-hand side of Eq. (\ref{chiq0}), $\chi_{jj}^{\mu\nu,\Omega^{2}}$, gives rise to a term with $\Omega^{2}\delta_{\mu\nu}$, which represents renormalization of the electric permittivity $\ez$ and we do not consider it further. 
The correlation function, Eq. (\ref{chiq}), is therefore dominated by the $\qv$-linear contribution with a coefficient  $g(i\Omega_{\ell})$, 
\begin{align}
\chi_{jj}^{\mu\nu}(\qv,i\Omega_{\ell})&\simeq - ig(i\Omega_{\ell})\sum_{\rho}\epsilon_{\mu\rho\nu}q_{\rho}. \label{chiql}
\end{align}

By rewriting the summation over the thermal frequency using the contour integral ($z\equiv i\omega_{n}$), Eq. (\ref{g}) becomes 
\begin{align}
g(i\Omega_{\ell})&=\frac{e^2\lambda^3}{24}\sum_{\kv}\sum_{\sigma_{1}\sigma_{2}\sigma_{3}}
\xi_{\kv, \sigma_{1}\sigma_{2}\sigma_{3}} \nnr
&\int_{C}\frac{{\rm d}z}{2\pi i}f(z){\rm g}_{\kv,\sigma_{1}}(z){\rm g}_{\kv,\sigma_{2}}(z) \nnr
&\times [{\rm g}_{\kv,\sigma_{3}}(z+i\Omega_{\ell})+{\rm g}_{\kv, \sigma_{3}}(z-i\Omega_{\ell})],
\end{align}
where $C$ is a counterclockwise contour surrounding the imaginary axis \cite{ADG75,AS06}, $f(z)\equiv(e^{\beta z}+1)^{-1}$ is the Fermi--Dirac distribution function, ${\rm g}_{\kv,\sigma}(z)\equiv \left[z-\epsilon_{\kv}^{\sigma}+i\eta{\rm sgn}({\rm Im} [z])\right]^{-1}$, and ${\rm Im}$ is the imaginary part. The retarded and advanced Green's function are defined as ${\rm g}_{\kv, \omega, \sigma}^{\rm r}\equiv{\rm g}_{\kv,\sigma}(\omega +i\delta)$ and ${\rm g}_{\kv, \omega, \sigma}^{\rm a}\equiv{\rm g}_{\kv,\sigma}(\omega -i\delta)$, respectively, where $\omega$ is an angular frequency of conduction electrons. Expanding $g$ with respect to $\Omega$ after the analytic continuation, the result up to the order of $\Omega^{2}$ reduces to
\begin{align}
g(\Omega +i\delta)\simeq g^{(0)}+i\Omega g^{(1)}+\Omega^{2}g^{(2)} ,\label{gomega}
\end{align}
with
\begin{align}
g^{(0)}&\equiv 
\frac{e^2\lambda^3}{6}\sum_{\kv,\omega}\sum_{\sigma_{1}\sigma_{2}\sigma_{3}}\xi_{\kv, \sigma_{1}\sigma_{2}\sigma_{3}}
f(\omega) \nnr
&\times {\rm Im}[{\rm g}_{\kv, \omega, \sigma_{1}}^{\rm r}{\rm g}_{\kv, \omega, \sigma_{2}}^{\rm r}{\rm g}_{\kv, \omega, \sigma_{3}}^{\rm r}],
\nnr
g^{(1)}&\equiv
\frac{e^2\lambda^3}{12}\sum_{\kv,\omega}\sum_{\sigma_{1}\sigma_{2}\sigma_{3}}\xi_{\kv, \sigma_{1}\sigma_{2}\sigma_{3}}
f'(\omega) \nnr
&\times {\rm Re}[{\rm g}_{\kv, \omega, \sigma_{1}}^{\rm r}{\rm g}_{\kv, \omega, \sigma_{2}}^{\rm r}({\rm g}_{\kv, \omega, \sigma_{3}}^{\rm r}-{\rm g}_{\kv, \omega, \sigma_{3}}^{\rm a})],
\nnr
g^{(2)}&\equiv -i
\frac{e^2\lambda^3}{24}\sum_{\kv,\omega}\sum_{\sigma_{1}\sigma_{2}\sigma_{3}}\xi_{\kv, \sigma_{1}\sigma_{2}\sigma_{3}} \nnr
&\times
i{\rm Im}
\begin{Bmatrix}
2f(\omega)
\begin{bmatrix}
{\rm g}_{\kv, \omega, \sigma_{1}}^{\rm r}{\rm g}_{\kv, \omega, \sigma_{2}}^{\rm r}{\rm g}_{\kv, \omega, \sigma_{3}}^{\rm r} \\
\times [({\rm g}_{\kv, \omega, \sigma_{1}}^{\rm r})^2 +({\rm g}_{\kv, \omega, \sigma_{2}}^{\rm r})^2 \\
+({\rm g}_{\kv, \omega, \sigma_{3}}^{\rm r})^2+{\rm g}_{\kv, \omega, \sigma_{1}}^{\rm r}{\rm g}_{\kv, \omega, \sigma_{2}}^{\rm r}]
\end{bmatrix}
\\
-f'(\omega)
\begin{bmatrix}
-{\rm g}_{\kv, \omega, \sigma_{1}}^{\rm r}{\rm g}_{\kv, \omega, \sigma_{2}}^{\rm r}({\rm g}_{\kv, \omega, \sigma_{3}}^{\rm a})^2
\\
+({\rm g}_{\kv, \omega, \sigma_{1}}^{\rm r})^2{\rm g}_{\kv, \omega, \sigma_{2}}^{\rm r}{\rm g}_{\kv, \omega, \sigma_{3}}^{\rm a}
\\
+{\rm g}_{\kv, \omega, \sigma_{1}}^{\rm r}({\rm g}_{\kv, \omega, \sigma_{2}}^{\rm r})^2{\rm g}_{\kv, \omega, \sigma_{3}}^{\rm a}
\end{bmatrix}
\end{Bmatrix} , \label{g0g1g2}
\end{align}
where 
${\rm g}_{\kv,\omega,\sigma}^{\rm r}\equiv(\omega-\epsilon_{\kv}^{\sigma}+i\eta)^{-1}$, ${\rm g}_{\kv,\omega,\sigma}^{\rm a}=({\rm g}_{\kv,\omega,\sigma}^{\rm r})^{\ast}$, ${\rm Re}$ is the real part, $f(\omega)$=$(e^{\beta \omega}+1)^{-1}$, and  $f'(\omega)\equiv\frac{\partial}{\partial\omega}f(\omega)$. The result can be simplified using the gauge invariance which impose  $g^{(0)}=0$. 
The second term on the right-hand side of Eq. (\ref{gomega}), the $\Omega$-linear term, generally arises from integrating out fermions coupled to bosons \cite{Caldeira81}, and gives rise to a term with $\Ev\cdot\Bv$, where $\Ev\equiv -\dot{\Av}$ and $\Bv\equiv\bm{\nabla}\times\Av$ are the electric and magnetic fields, respectively. 
However, we drop the term because $\Ev\cdot\Bv=\frac{1}{4}\sum_{\mu\nu\alpha\beta}\epsilon^{\mu\nu\alpha\beta}F_{\mu\nu}F_{\alpha\beta}$ reduces to a surface term by the divergence theorem, where $\epsilon^{\mu\nu\alpha\beta}$ is a totally antisymmetric tensor,  $F_{\mu\nu}\equiv\partial_{\mu}A_{\nu}-\partial_{\nu}A_{\mu}$ is the field strength of the electromagnetic field. 
We thus have $g\sim\Omega^{2}g^{(2)}$ and the effective Hamiltonian finally turns out to be Eq. (\ref{Hop}) with $g=\frac{g^{(2)}}{4}$.
Explicit evaluation of the coefficient $g^{(2)}$ is carried out in the case of free electron limit neglecting the spin polarization in  Appendix\ref{B}. 

%%%%%%%%%%%%%%%%%%%%%%%%%%%%%%%%%%%%%%%%%%%%%%%%%%%%%%%%%%%%%%%%%%%%%%%%%%%%%%%%%%%%%%%%%%%
\section{Summary} 
Using an imaginary-time path-integral formalism, we derived an effective Hamiltonian of the electromagnetic fields in terms of an optical chirality order parameter in a Weyl spin-orbit system having quadratic dispersion. The effective Hamiltonian approach clearly revealed that natural optical activity in the system is due to the emergence of the optical chirality order parameter. 

%%%%%%%%%%%%%%%%%%%%%%%%%%%%%%%%%%%%%%%%%%%%%%%%%%%%%%%%
\section*{Acknowledgments}
The authors thank A. Shitade for valuable discussions.
This work was supported by a Grant-in-Aid for Scientific Research (B) (No.17H02929) from the Japan
Society for the Promotion of Science and a Grant-in-Aid for Scientific Research on Innovative Areas (No.26103006) from
The Ministry of Education, Culture, Sports, Science and Technology (MEXT), Japan.
H. K. was supported by RIKEN Junior Associate Program.

%%%%%%%%%%%%%%%%%%%%%%%%%%%%%%%%%%%%%%%%%%%%%%%%%%%%%%%%%%%%%%%%%%%%%%%%%%%%%%%%%%%%%%%%%%%
\appendix
\section{Derivation of Eqs. (\ref{chiq}), (\ref{g}), and (\ref{xi})} \label{A}
This section shows the detailed derivation of Eqs. (\ref{chiq}), (\ref{g}), and (\ref{xi}) from Eq. (\ref{chi}). 
By use of 
\begin{align}
{\rm tr}[\sigma_{\alpha}\sigma_{\beta}\sigma_{\gamma}\sigma_{\delta}]&=
2(\delta_{\alpha\beta}\delta_{\gamma\delta}+\delta_{\beta\gamma}\delta_{\alpha\delta}-\delta_{\alpha\gamma}\delta_{\beta\delta}), \nnr 
{\rm tr}[\sigma_{\alpha}\sigma_{i}\sigma_{\beta}\sigma_{j}\sigma_{\gamma}]&=
2i(\delta_{\alpha i}\epsilon_{\beta j \gamma}+\delta_{j\gamma}\epsilon_{\alpha i \beta}+\delta_{\beta j}\epsilon_{\alpha i\gamma}-\delta_{\beta\gamma}\epsilon_{\alpha ij}),
\end{align}
the result up to the linear order in $\qv$ reads
\begin{align}
\chi_{jj}^{\mu\nu}(\qv,i\Omega_{\ell})&\simeq \chi_{jj}^{\mu\nu}(\qv=0,i\Omega_{\ell})
-\frac{(-i)e^2\lambda^2}{\beta V}\sum_{n,\kv}\sum_{\rho} \nnr
&\times
\begin{Bmatrix}
&-{\displaystyle \sum_{i}}\frac{1}{\lambda m}\gamma_{\qv}^{\rho}\hat{\gamma}_{\kv}^{i}(\epsilon_{\mu i\rho}k_{\nu}-k_{\mu}\epsilon_{\nu i \rho})
(h_{\kv,n}^2-j_{\kv,n}^2)(j_{\kv,n+\ell}+j_{\kv,n-\ell}) \nnr
&-\frac{k_{\rho}q_{\rho}}{m}{\displaystyle \sum_{i}}\hat{\gamma}_{\kv}^{i}\epsilon_{\mu\nu i}
\begin{bmatrix}
(h_{\kv,n}^2+j_{\kv,n}^2)(j_{\kv,n+\ell}+j_{\kv,n-\ell}) \nnr
-2h_{\kv,n}j_{\kv,n}(h_{\kv,n+\ell}+h_{\kv,n-\ell})
\end{bmatrix}
\nnr 
&+\gamma_{\qv}^{\rho}
\begin{bmatrix}
\epsilon_{\mu\nu\rho}(h_{\kv,n}^2+j_{\kv,n}^2)(h_{\kv,n+\ell}+h_{\kv,n-\ell}) \\
+2j_{\kv,n}^2{\displaystyle \sum_{mn'op}}\epsilon_{\rho mn'}\epsilon_{n'op}
\epsilon_{\mu\nu p}\hat{\gamma}_{\kv}^{m}\hat{\gamma}_{\kv}^{o}
(h_{\kv,n+\ell}+h_{\kv,n-\ell}) \\
+2{\displaystyle \sum_{i}}\hat{\gamma}_{\kv}^{i}\hat{\gamma}_{\kv}^{\rho}\epsilon_{\mu i \nu}h_{\kv,n}j_{\kv,n}(j_{\kv,n+\ell}+j_{\kv,n-\ell})
\end{bmatrix} 
\end{Bmatrix} ,\label{chiex} \\
\end{align}
where $\bm{\hat{\gamma}}_{\kv}\equiv \bm{\gamma}_{\kv}/|\bm{\gamma}_{\kv}|$ and $\bm{\gamma}_{\qv}\equiv -\lambda\qv$. Here  $h_{\kv,n}$ and $j_{\kv,n}$ are defined respectively as 
\begin{align}
h_{\kv,n}&\equiv\frac{1}{2}\sum_{\sigma=\pm}{\rm g}_{\kv,n,\sigma} ,\nnr 
j_{\kv,n}&\equiv\frac{1}{2}\sum_{\sigma=\pm}\sigma {\rm g}_{\kv,n,\sigma} \label{hj},
\end{align}
and $\sigma$ = $\pm$ is the diagonalized spin index. 
Using a rotational symmetry in $k$-space, i.e., $k_{\mu}k_{\nu}=\frac{\kv^2}{3}\delta_{\mu\nu}$ and Eq. (\ref{hj}), the result of Eq. (\ref{chiex}) is summarized in Eqs. (\ref{chiq}), (\ref{g}), and (\ref{xi}).

\section{Calculation of $g^{(2)}$} \label{B}
We here calculate $g^{(2)}$ shown in Eq. (\ref{g0g1g2}) in the free electron limit. 
Performing an integration by parts with respect to $\omega$ by use of ${\rm g}_{\kv,\omega}^{\rm r}\equiv(\omega-\epsilon_{\kv}+i\eta)^{-1}$, $g^{(2)}$ reduces to
\begin{align}
g^{(2)} &=-\frac{e^2\lambda^3}{2\pi} \sum_{\kv} {\rm Im}
\biggl[
2({\rm g}_{\kv}^{\rm r})^4 + ({\rm g}_{\kv}^{\rm r}{\rm g}_{\kv}^{\rm a})^2-2({\rm g}_{\kv}^{\rm r})^3 {\rm g}_{\kv}^{\rm a}
\biggl],
\end{align}
where ${\rm g}_{\kv}^{\rm r}\equiv{\rm g}_{\kv,\omega =0}^{\rm r}$. 
Carrying out the integral over $k$ in the case of $\epsilon_{\rm F}\tau_{\rm e}\gg1$, 
the dominant contribution turns to be 
\begin{align}
g^{(2)} \simeq -12(\pi \epsilon_{\rm F}\tau_{\rm e})^2 \frac{e^2\lambda^3 \nu }{\epsilon_{\rm F}^3}\tau_{\rm e}^2
,
\end{align}
where $\nu \equiv\frac{m^{\frac{3}{2}}}{\sqrt{2}\pi^2}\sqrt{\epsilon_{\rm F}}$ is the density of states at the Fermi energy per unit volume. 

%%%%%%%%%%%%%%%%%%%%%%%%%%%%%%%%%%%%%%%%%%%%%%%%%%%%%%%%
\bibliographystyle{jplain}
\bibliography{180327temp}

%%%%%%%%%%%%%%%%%%%%%%%%%%%%%%%%%%%%%%%%%%%%%%%%%%%%%%%%

\end{document}